\documentstyle[12pt]{article}
\def\beq{\begin{equation}}
\def\eeq{\end{equation}}
\newcommand{\bea}{\begin{eqnarray}}
\newcommand{\eea}{\end{eqnarray}}
\newcommand{\nn}{\nonumber}
\def\noi{\noindent}
\font\boldgreek=cmmib10
\mathchardef\mysigma="091B
\def\bfsigma{{\fam=9 \mysigma}\fam=1}
\mathchardef\myalpha="090B
\def\bfalpha{{\fam=9 \myalpha}\fam=1}
\begin{document}
\vbox to 3.5 truecm{}
\begin{flushright}
LPTHE-Orsay 95/11,\\
hep-ph/9504267
\end{flushright}
\begin{center}
{\large \bf WEAK TRANSITIONS OF HEAVY MESONS} \\
{\large \bf AND THE QUARK MODEL} \\
\vspace{6.0 ex}
{\large A. Le Yaouanc, L. Oliver, O. P\`ene and J.-C. Raynal} \\
\vspace{2.5ex}
{\large Presented by L. Oliver}\\
\vspace{4.0 ex}
{\large \it Laboratoire de Physique Th\'eorique et Hautes
Energies\footnote{Laboratoire associ\'e au
Centre National de la Recherche Scientifique - URA 63}} \\
{\large \it Universit\'e de Paris XI, B\^atiment 211, 91405 Orsay Cedex,
France} \\
\vspace{3.5ex}
{\large Talk delivered at the Journ\'ees sur les projets \\
de Physique Hadronique, Soci\'et\'e Fran\c caise de Physique,\\
Super-Besse (France), 12-14 janvier 1995} \\
\end{center}
\vspace{4.5ex}
\noi {\bf R\'esum\'e}
\vspace{3.0ex}

Nous soulignons les caract\'eristiques du mod\`ele des quarks et ses succ\`es
dans la
description du spectre et des transitions de hadrons lourds ou l\'egers,
comparant bri\`evement aux
premiers principes de QCD. Pour montrer l'utilit\'e actuelle des techniques et
intuition physique du
mod\`ele, nous pr\'esentons un mod\`ele des quarks semi-relativiste des
facteurs de forme des m\'esons qui
pr\'esente l'invariance d'\'echelle d'Isgur-Wise, et des corrections \`a
celle-ci. Comme exemples
d'applications, nous consid\'erons les facteurs de forme semi-leptoniques et
des d\'esint\'egrations
non leptoniques de m\'esons \`a saveur lourde. Nous posons la question de la
violation de la
factorisation, et montrons comment le mod\`ele des quarks peut donner des
indications sur ce
probl\`eme.
\vskip 1 truecm
\noi {\bf Abstract}
\vspace{3.0ex}

We underline the general features of the constituent quark model and its
success in the description of
the spectrum and transitions of hadrons made up of light or heavy quarks,
briefly compared to first
principles of QCD. To show the present usefulness of the techniques and
intuitive insight of the
model we present a semirelativistic quark model of meson form factors that
exhibits Isgur-Wise
scaling and corrections to it. As examples of applications, we consider
semileptonic form factors and
non-leptonic decays of heavy flavor mesons. We address the issue of the
violation of factorization,
and show how the quark model can give hints on the problem.

\vspace{5.0ex}
\section{\bf Success and limits of the quark model} \par

Following earlier remarks$^1$, we will first briefly describe the main
characteristics of the Constituent Quark Model (CQM), comparing it to the first
principles of QCD,
that we believe to be the right theory of strong interactions. We underline the
relations and
differences between both approaches, in particular which aspects of the CQM are
understood in QCD,
and also the successes of the CQM that still wait for rigorous explanations.
\par

1) The CQM assumes a fixed number of constituents in a given hadron.  \par

On the contrary, in QCD the number of constituents is not fixed, since
$\alpha_s(Q^2)$ grows with the
distance, and in particular the number of gluons and quarks in the hadron wave
function depends on
$Q^2$. \par

2) In the CQM one considers constituents masses~; for the light quarks these
are of the order $m \simeq $
0.3 GeV (from $\mu_p = \displaystyle{1 \over 2m} = \displaystyle{2.79 \over
2M_N}$). The constituent
masses, together with a non-relativistic approximation, results in an
approximate SU(6) symmetry for
the light hadrons, non-strange and strange, that works reasonably well, mainly
in the baryon sector.
\par

In QCD, one starts with current masses in the Lagrangian, masses that for the
lightest quarks $u$,
$d$ are of the order of a few MeV, resulting in an approximate Chiral Symmetry,
that is dynamically
broken. The constituent mass of the quark model can be identified with the
quark mass dynamically
generated by this spontaneous chiral symmetry breaking. However, the
approximate SU(6) symmetry of the
CQM, although empirically successful, does not have a theoretical basis. One
must recall that
attempts have been made to include the phenomenon of dynamical symmetry
breaking in quark models,
including dynamically generated mass, quasi-Goldstone mesons and degeneracy of
the vacuum. These
are introduced thanks to a second quantized chiral-symmetric version of the
relativistic potential.
But these attempts have not yet reached the stage of a complete phenomenology
comparable to the CQM.
\par

3) In the CQM one assumes a Schr\"odinger equation with a flavor-independent
confining potential with
a short distance and a long distance pieces~:
\beq
V(r) = - {\kappa \over r} + \lambda  r + C
\eeq

\noi (for heavy quarkonia, $\kappa = 0.5$, $\lambda \simeq 0.2$ GeV$^2$).
Empirically, one finds that a
Lorentz vector short distance, that in QCD would correspond to one-gluon
exchange (OGE), and a
Lorentz Scalar long distance pieces are favored, as regards spin-dependent
forces. Indeed, within
this hypothesis, one has $v^2/c^2$ spin-dependent corrections to the spectrum,
a short distance
spin-spin interaction interaction,
\beq
V_{SS}({\bf r}) = {{\bfsigma}_1\cdot {\bfsigma}_2 \over 6m_1m_2} \Delta
V_V({\bf r}) \Rightarrow
{16\pi \alpha_s \over 3} \  {{\bfsigma}_1 \cdot {\bfsigma}_2 \over 6m_1m_2}
\delta({\bf r}) \quad
\hbox{(OGE)}	 \eeq

\noi a ${\bf L} \cdot {\bf S}$ force
\beq
V_{LS}({\bf r}) = {{\bf L}\cdot {\bf S} \over 2m^2r} \left ( 3 {dV_V \over dr}
- {dV_S \over dr}
\right )  \quad (m_1 = m_2)
\eeq

\noi and a tensor one
\beq
V_T({\bf r}) = {S_{12} \over 12m_1m_2} \left ( {1 \over r} \ {dV_V \over dr} -
{d^2V_V \over dr^2}
\right ) \ \ \ .
\eeq

The ratio of level differences in quarkonium would be, in the hypothesis of
pure Lorentz vector
potential (with $\beta = <\displaystyle{\lambda \over
r}>/<\displaystyle{\alpha_s \over r^3}>$)~:
\beq
R = {M(\chi_2) - M(\chi_1) \over M(\chi_1) - M(\chi_0)} = {4 \over 5} \  {8 +
7\beta \over 8 +
4\beta} > {4 \over 5}      \quad  \hbox{(Lorentz \ Vector)} \eeq

\noi while, assuming the short distance potential to be Lorentz vector and the
long distance one
Lorentz scalar~:
\beq
R = {4 \over 5} \  {16 - 5\beta \over 16 - 2\beta}  <  {4 \over 5}  \qquad
\hbox{(SD \ Vector, \ LD \
Scalar)}	\ \ \ . \eeq

Experiment gives for the $c\bar{c}$ system $R(\chi_c) = 0.48$ while for the
$b\bar{b}$ one
$R(\chi_b) = 0.64$, $R(\chi '_b) = 0.58$, strongly pointing to the spin-orbit
force being like for a
Lorentz scalar. Moreover, one has $(M_{cog}(^{3}P_J)$, $M(^{1}P_1$) are
independent of $V_T$)~:
 \beq
M_{cog}(^{3}P_J) = M(^{1}P_1)
\eeq

\noi that is very well verified in charmonium (Table 2), where $M_{cog}$ is the
average mass of
$\chi_0$, $\chi_1$, $\chi_2$. It is important to emphasize that there is no
long range spin-spin
force, as it would be if the long distance potential were a Lorentz vector.\par

These empirical hints of the quark model are well justified in QCD. Indeed,
confinement is deduced
from lattice QCD at strong coupling$^{2}$. More realistic calculations on the
lattice even
with dynamical fermions suggest a potential of the form (1), as we can see in
Fig. 1$^{3}$.
Moreover, Wilson loop calculations with an expansion in $1/m_Q$ confirm the
Schr\"odinger picture with
the potential (1) and the form of the ${\bf S} \cdot {\bf S}$ and ${\bf L}
\cdot {\bf S}$ interactions
as suggested by the quark model$^{4}$. \par

4) In the CQM the vacuum is trivial. In QCD the vacuum is highly non-trivial.
One has three main sets
of phenomena~: (i) Non perturbative effects connected with confinement: $<GG>$
condensate.
(ii) Dynamical breaking of chiral symmetry, $<q\bar{q}>$ condensate,
quasi Nambu-Goldstone bosons $\pi$ and $K$. (iii) Non-perturbative effects in
the U(1) sector : $<G\tilde G>$
condensate, non-triviality of the vacuum due to instantons, special status of
the $\eta '$
pseudoscalar meson, possibility of CP violation in the strong interactions.
Concerning (ii), as
pointed out above, one can formulate generalizations of the quark model with
spontaneous chiral
symmetry breaking$^5$.

5) The Quark Model is a non-relativistic model, the expansion parameter being
$v/c$. The mean value of
this parameter is quite different in the different bound systems~:
\begin{center}
Table I \\
 Quark mass, level spacing $\omega$ and internal $v^2/c^2$ for the different
quarkonia \\
\vskip 3 mm
\begin{tabular}{|c|c|c|c|} \hline
Quarkonia	&$m_q$ (GeV)	&$\omega$ (GeV)	&$v^2/c^2$ \\ \hline
& & & \\
$b\bar{b}$ 	&5.12	&0.48	&0.13 \\ \hline
$c\bar{c}$  &1.82	&0.46	&0.27 \\ \hline
$q\bar{q}$ 	&0.3	&0.5-0.7	&0.7 \\ \hline
$Q\bar{q}$	&0.3	&0.46	&0.7 \\ \hline
\end{tabular}
\end{center}

\noi For the light quarks the internal quark motion is relativistic. In the
quark model one considers
two types of relativistic effects : (i) Relativistic corrections due to the
binding, for example
in current matrix elements like the nucleon axial coupling
\beq
{G_A \over G_V} = {5 \over 3} \  (1 - 2\delta)
\eeq

\noi where $\delta = O\left ( {v^2 \over c^2} \right )$, has the right sign and
order of magnitude (there are also $O(\alpha_s)$ radiative corrections). (ii)
Relativistic effects due to the center-of-mass motion, that come out in
hadron form factors for example, as we will see below. \par

In QCD, there are three important limiting regimes around which one can
formulate very fruitful systematic expansions in small parameters: (i) The hard
or large momentum  regime, in which the expansion parameter is
$\alpha_s(Q^2) \sim 1/\log (Q^2/\Lambda^2)$.  (ii) Chiral symmetry limit for
light
quarks  where the expansion parameters are, grosso modo, $m_q/\Lambda$ and
$p_\mu/\Lambda$ (small momenta). Chiral symmetry translates into effective
light hadron chiral Lagrangians.  (iii) Heavy Quark limit where the expansion
parameter is essentially $\Lambda/m_Q$. This leads to the useful Heavy Quark
Symmetry.\par

	As we will see below, the relativistic effects due to the center-of-mass
motion in the CQM make the
link with the Heavy Quark Effective Theory of QCD. \par

	6) The CQM gives a good overall description of spectra and transitions of
light and heavy hadrons,
for the ground state and the excitations. We show a few examples in Fig. 2
(energy levels of first
excited light baryons), Table 2 (energy levels of heavy quarkonia $c\bar{c}$,
$b\bar{b}$), Table 3
(radiative transitions of baryon isobars) and Table 4 (E1 transitions in the
$b\bar{b}$ system).
However, one must say that for the quasi-Goldstone bosons, mainly the pion, and
the U(1) sector, the
quark model finds difficulties. \par

	The quark model provides an intuitive understanding of the phenomena in terms
of bound states and
its wave functions, that is used almost unconsciously by everybody working on
phenomena involving
hadrons. \par

	The rigorous methods of QCD, like lattice QCD, confirm many results of the
quark model for the
ground state hadrons, and in addition can treat the pion as a Goldstone boson.
However, these methods
cannot yet give an overall view of the spectra and transitions of hadrons as
the CQM does,
especially because it handles very easily the excited states, which are hardly
accessible to the
fundamental methods. Moreover, the success of the non-relativistic quark model
for light quarks,
even if it is possibly amended by relativistic corrections, is not understood
in terms of QCD.

\begin{center} Table 2 \\ Quarkonia $c\bar{c}$, $b\bar{b}$ levels ($NR$~:
non-relativistic model~;
${\bf S} \cdot {\bf S}$~: hyperfine splitting)~; *~: input \\
\vskip 3 mm
\begin{tabular} {|c|c|c|c|c|c|} \hline
& & & & & \\
$c\bar{c}$	&Exp. (GeV)	&$NR [6] {\bf S}\cdot {\bf S} [7]$ 	&$b\bar{b}$	 &Exp.
(GeV)	&$NR [6] {\bf S}\cdot
{\bf S} [7]$ \\ \hline
$\eta_c$  $1^1S_0$ &2.980	&2.956	&$\eta_b$ $1^1S_0$		&  &9.30 \\ \hline
$J/\psi$  $1^3 S_1$ 	&3.097	&3.095*	&$\Upsilon$ $1^3S_1$ &9.460	&9.46* \\
\hline
$\eta_c$ $2^1S_0$ &3.594	&3.550	&$\eta_b$ $2^1S_0$	&	&9.93 \\ \hline
$\psi$ $2^3S_1$ &3.686	&3.684*	&$\Upsilon$ $2^3S_1$ &10.023	&10.05 \\ \hline
$\psi$ $3^3S_1$ & 4.040	&4.110 &$\Upsilon$ $3^3S_1$ &10.355	&10.40 \\ \hline
$\psi$ $4^3S_1$ &4.415	&4.460	&$\Upsilon$ $4^3S_1$ &10.580	&10.67 \\ \hline
& &	& $\eta_b$ $4^1S_0$		&  &10.57 \\ \hline
& & & $\Upsilon$ $5^3S_1$ &10.865	&10.92 \\ \hline
& & & $\eta_b$ $5^1S_0$	&	&10.82 \\ \hline
&	&	& $\Upsilon$ $6^3S_1$ &11.019 & \\ \hline
& & & $\eta_b$ $6^1S_0$	&	 &11.04 \\ \hline
$\chi_c$ $1^3P_J$ &3.525	&3.522* 	&$\chi_b$ $1^3P_J$ &9.900	&9.96  \\
(c.o.g.) & & &(c.o.g.) & & \\ \hline
$h_c$ $1^1P_1$ &3.526	& &$h_b$ $1^1P_1$	 & &9.96  \\ \hline
$\psi$ $1^3D_1$ &3.770	& 3.810	&$\Upsilon$ $1^3D_1$ & &10.20  \\ \hline
& & &$\chi_b$ $2^3P_J$ &10.261 &10.31 \\
& & &(c.o.g.) & & \\ \hline
& & &$h_b$ $2^1P_1$	 & &10.31 \\ \hline
$\psi$ $2^3D_1$ &4.159	&4.190	&$\Upsilon$ $2^3D_1$ & &10.50 \\ \hline
\end{tabular}
\end{center}

\newpage

\begin{center}
Table 3 \\
Radiative transitions $N^* \to N\gamma$ (data from PDG 1994). \\
\vskip 3 mm
\begin{tabular}{|c|c|c|c|c|c|c|c|c|} \hline
&$A^p_{3/2}$ &$A^p_{3/2}$ &$A^p_{1/2}$ &$A^p_{1/2}$ &$A^n_{3/2}$ &$A^n_{3/2}$
&$A^n_{1/2}$
&$A^n_{1/2}$ \\
&th.	&exp.	&th.	&exp. &th.	&exp.	 &th.	&exp. \\ \hline
{\bf 56},0$^+$ & & & & & & & & \\
$P_{33}$ &- 178	&- 257	&- 103	&- 141 & & & & \\
(1236) & & & & & & & & \\ \hline
{\bf 70},1$^-$ & & & & & & & & \\
$S_{11}$ & & &160	&68	&	&	&- 109	&- 59 \\
(1535) & & & & & & & & \\ \hline
{\bf 70},1$^-$ & & & & & & & & \\
$D_{13}$ &112	&163	&- 29	&- 22 &- 112	&- 137	&- 30	&- 62 \\
(1520) & & & & & & & & \\ \hline
{\bf 70},1$^-$ & & & & & & & & \\
$D_{15}$ &0	&18	&0	&18	&- 53	&- 70	&- 38	&- 50 \\
(1670) & & & & & & & & \\ \hline
{\bf 70},1$^-$ & & & & & & & & \\
$S_{31}$ & & &47	&30 & & & & \\
(1650) & & & & & & & & \\ \hline
{\bf 70},1$^-$ & & & & & & & & \\
$D_{33}$ &91	&91	&92	&114	& & & & \\
(1670)	& & & & & & & & \\ \hline
{\bf 56},2$^+$ & & & & & & & & \\
$F_{15}$ &70 &135 &- 15	&- 14	&0	&- 35	&41	&27 \\
(1688) & & & & & & & & \\ \hline
\end{tabular}
\end{center}

\begin{center}
Table 4 \\
E1 radiative transitions in the $b\bar{b}$ system. \\
The ratio $\Gamma / k^3 \propto (2J + 1)$, predicted by the Non-relativistic
Quark Model, is very well satisfied by the data. \\

\vskip 3 mm
\begin{tabular} {|c|c|c|c|c|c|} \hline
$\Upsilon (2S) \to$ &$\Gamma (keV)$ &$\Gamma (keV)$ &$\Upsilon (3S) \to$
&$\Gamma (keV)$
&$\Gamma (keV)$ \\
$\chi_b(1^3P_J)$ $\gamma$ &exp.	&th.$^8$	&$\chi_b (2^3P_J)$ $\gamma$	&exp.
&th.$^8$ \\ \hline
$J = 0$	&1.9	&1.43	&$J = 0$	&1.5	&1.55 \\ \hline
$J = 1$	&2.95	&2.27	&$J = 1$	&2.8	&2.5 \\ \hline
$J = 2$	&2.90 &2.24	&$J = 2$	&2.7	&2.75 \\ \hline
\end{tabular}
\end{center}

\vskip 5 mm
\section{\bf Weak transitions of heavy mesons}

Weak decays of mesons are governed by the Cabibbo-Kobayashi-Maskawa (CKM)
matrix~:
\beq
(\bar{u},  \bar{c},  \bar{t}) \gamma_{\mu}(1 - \gamma_5) \left (
\begin{array}{ccc}
V_{ud}  &V_{us} &V_{ub} \\
V_{cd}  &V_{cs}  &V_{cb} \\
V_{td}  &V_{ts}  &V_{tb}
\end{array} \right ) \left ( \begin{array}{c}
d \\
s \\
b
\end{array} \right )
\eeq

\noi that, in the Wolfenstein parametrization and expansion in powers of the
Cabibbo angle $\lambda =
V_{us} = \sin \theta_C$, writes~:

\beq
V \cong  \left ( \begin{array}{ccc}
1 - {\lambda^2 \over 2}  &\lambda  &A\lambda^3(\rho - i\eta)\\
- \lambda &1 - {\lambda \over 2} &A\lambda^2\\
A\lambda^3(1 - \rho -i \eta) &-A\lambda^2 &1
\end{array} \right ) \ \ \ .
\eeq

\noi In this expansion, unitarity is approximate, $A$ is of order $O(1)$, and
the complex number
$\rho -i\eta = e^{-i\delta}$ is responsible of the CP violation, with $\rho$
and $\eta$ of order
$O(1)$. \par

The different decays depend on different CKM matrix elements. For example, the
semileptonic decays $B \to
D(D^*) \ell \nu$, $B \to \pi (\rho ) \ell \nu$, $D \to K(K^*) \ell \nu$, $K \to
\pi \ell \nu$ depend
respectively on $V_{cb}$, $V_{ub}$, $V_{cs}$ and $V_{us}$. \par \vskip 5 mm
\subsection{Heavy Quark Symmetry}

Assume a $Q\bar{q}$ system made of a heavy quark $Q$ and a light antiquark
$\bar{q}$. As we have seen
above, in the Quark Model, the system is described by a Hamiltonian with a spin
independent potential
plus spin dependent terms that are inversely proportional to the quark masses~:

\beq
	H = {p^2 \over 2\mu} + V(r) +  {16\pi \alpha_s \over 3} {\bfsigma_1 \cdot
\bfsigma_2 \over 6m_Qm}
\delta ({\bf r}) + \cdots	 \qquad \mu = {m_Qm \over m_Q+m} \ \ \ .		 \eeq

\noi In the infinitely heavy quark limit, for $m_Q \to \infty$, $H \to$ finite
limit~: the reduced
mass $\mu \to m$, the wave functions, binding energies, become independent of
the flavor and spin of
the heavy quark. The dynamics depends only then on the light quark degrees of
freedom~:
$H({\bf r},{\bf p},\bfsigma_2,m)$. This has a number of interesting
consequences for hyperfine and
fine splittings~:

\bea
&&{\bf S} \cdot {\bf S} \  {\rm splittings} :	\qquad	M_{B^*} = M_B	\quad
M_{D^*} = M_D \nn \\
&&{\bf L} \cdot {\bf S} \ {\rm splittings} :  \qquad		M_{B^{**}} - M_B =
M_{D^{**}} - M_D
\eea

\noi that are qualitatively satisfied by experiment, with small corrections to
the infinitely heavy
quark limit. \par

One can generalize to QCD these simple considerations, with an exact treatment
of light degrees of
freedom, gluons and light quarks$^{9}$. For an arbitrary four-velocity of the
heavy
hadron $v^{\mu} = P^{\mu}/M$ $(v^2 = 1)$, one can write the four-momentum of
the heavy quark in the
form

\beq
p^{\mu}_Q = m_Qv^{\mu} + k^{\mu}
\eeq

\noi where $k$ is a typical residual momentum due to the binding within the
heavy-light hadron wave
function. Then, the heavy quark propagator can be approximated by

\beq
{m_Q{/\hskip - 2 mm v} + {\ \hskip - 2 mm k} + m_Q \over \left ( m_Qv + k
\right )^2 - m^2_Q}  \cong
{{/\hskip - 2 mm v} + 1 \over 2v \cdot k}	\ \ \ .			 \eeq

\noi The interesting feature of this expression is that the heavy propagator is
independent of the
heavy mass and depends only on the hadron four-velocity and a residual momentum
$k$, very small
compared to $m_Qv$. \par

One has a generalization of an atomic picture. Like in the hydrogen atom, if
the hadron is at rest,
the heavy quark acts as a static source of color (up to $1/m_Q$ corrections).
However, here the
picture is fully relativistic, since it is generalized to any hadron
four-velocity, and this will
have interesting consequences for the hadron form factors. One has an
independence of the spin and
flavor of the heavy quark, leading to a symmetry $[SU(2N_f)]_v$, the subindex
$v$ meaning that there
is a symmetry for any given $v^{\mu}$. \par

This Heavy Quark Symmetry, discovered by Isgur and Wise$^{9}$, has important
consequences for the
semileptonic form factors in the heavy-to-heavy case (like $B \to D(D^*) \ell
\nu$) and also, although
weaker results, for the heavy-to-light case (like $B \to \pi(\rho) \ell
\nu$)$^{10}$. \par

	Let us define the form factors :

\bea
&&< P_i|V_{\mu}|P_j> = \left ( p^i_{\mu} + p^j_{\mu} -{M^2_j - M^2_i \over q^2}
 q_{\mu} \right ) f_+
(q^2) + {M^2_j - M^2_i \over q^2}  q_{\mu} f_0(q^2) \nn \\
&&< V_i|A_{\mu}|P_j > = \left ( M_i + M_j \right ) A_1(q^2)
\left ( \varepsilon^*_{\mu} - {\varepsilon^* \cdot q \over q^2}  q_{\mu} \right
) - \nn \\
&&- A_2(q^2) {\varepsilon^*\cdot q \over M_i + M_j} \left ( p^i_{\mu} +
p^j_{\mu} - {M^2_j - M^2_i
\over q^2}  q_{\mu} \right ) + 2 M_i A_0(q^2)  {\varepsilon^* \cdot q \over
q^2}  q_{\mu} \nn \\
&&<V_i|V_{\mu}|P_j > = i {2V(q^2) \over M_i + M_j} \  \varepsilon_{\mu \nu \rho
\sigma} \
p^{\nu}_j \ p^{\rho}_i \ \varepsilon^{*\sigma} 	\ \ \ .
\eea

\noi The Heavy Quark Symmetry implies scaling laws for the heavy-to-light form
factors$^{10}$, that
apply for example to the form factors $B \to \pi(\rho)\ell \nu$ and also to $B
\to K(K^*)$, that could
be related to the non-leptonic decays $B \to K(K^*)\psi$ within the
factorization approximation.
The scaling law applies at fixed three-momentum ${\bf q}$, small compared to
the heavy quark mass,
i.e. four momentum transfer close to its maximum value~:

\beq
{1 \over \sqrt{m_Q}} < K(K^*),{\bf q}|J^{\mu}|P_Q > \cong  Cte.     \qquad
(|{\bf q}|, M_K, M_{K^*}
\ll  m_Q)	 \ \ \ . \eeq

\noi This implies, for the different form factors~:

\bea
&&f_0({\bf q}), A_1({\bf q}) \sim  {1 \over \sqrt{m_Q}} \nn \\
&&f_+({\bf q}), V({\bf q}), A_0({\bf q}), A_2({\bf q}) \sim \sqrt{m_Q} 	\quad
\left ( |{\bf q}| \ll
M_Q , q^2 \cong q^2_{max} \right ) \ \ \ .
\eea

\noi In the heavy-to-heavy case, like in $B \to D(D^*)\ell \nu$, stronger
relations are valid for any
value of $q^2$ (up to corrections in $1/m_Q$, in practice $1/m_c$)$^{9}$~:

\bea
&&{\sqrt{4M_B M_{D^*}} \over (M_B + M_{D^*})} A_0(q^2) = {\sqrt{4M_B M_{D^*}}
\over (M_B + M_{D^*})}
A_2(q^2) = {\sqrt{4M_B M_{D^*}} \over (M_B + M_{D^*})} {A_1(q^2) \over \left [
1 - {q^2 \over (M_B +
M_{D^*})^2} \right ]} = \nn \\
&&(\sqrt{4M_B M_D} \over (M_B + M_D)} f_+(q^2) = {\sqrt{4M_B M_D} \over (M_B +
M_D)} {f_0(q^2) \over
\left [ 1 - {q^2 \over (M_B + M_D)^2} \right ]} = {\sqrt{4M_B M_{D^*} \over
(M_B + M_{D^*})} V(q^2)
= \xi (v \cdot v') \nn \\
\eea

\noi where $\xi (v\cdot v')$ is the so-called Isgur-Wise function that depends
only on the product of
the initial and final four-velocities. It satisfies the normalization $\xi (1)
= 1$ (i.e. at
${\bf q} = 0$ or $q^2 = q^2_{max})$. \par

	These relations and the normalization are very important because they allow an
almost model-independent determination of the CKM matrix element $V_{cb}$ from
the decay $B \to D^*\ell \nu$. \par

	The differential decay rate $B \to D^*\ell \nu$ writes

	\beq
{d\Gamma \over dw} = {G^2 \over 48\pi^3}  F(M_{D^*},M_B,w) |V_{cb}|^2  \
\eta^2_A
\ \widehat{\xi}^2(w)
\eeq

\noi where $w$ is the variable $w = v_{D^*} \cdot v_B = {M^2_B + M^2_{D^*} -q^2
\over 2M_BM_{D^*}}$,
$F(M_{D^*},M_B,w)$ is a known function, $\eta_A$ is a known short distance QCD
correction, and the
function $\widehat{\xi}(w)$ differs from the exact heavy quark limit function
$\xi (w)$ by $O(1/m_Q)$
corrections. In particular, at the normalization point, there are quadratic
corrections$^{11}$~:
$\widehat{\xi}(1) = 1 + \delta_{1/m^2}$. From a linear fit $\widehat{\xi}(w) =
1 -
\widehat{\rho}^2(w-1)$ one finds (CLEO experiment$^{12}$)

\beq
|V_{cb}| = 0.038 \pm 0.006 \pm 0.004	   	\ \ \ .
\eeq
\vskip 5 mm
\subsection{Quark model of semileptonic form factors}

This is a weak binding model that takes into account the relativistic
center-of-mass motion of the hadrons$^{13}$,
as refered to in 5) of section 1. The model provides an intuitive link
between the Quark Model and the Heavy Quark Symmetry. \par

	The total hadron wave function writes
\beq
\Psi^{tot}_{\bf P}(\{{\bf p}_i\}) = \delta (\sum_i {\bf p}_i - {\bf P})
\Psi_{\bf P}(\{{\bf
p}_i\})
\eeq

\noi with the internal wave function
\beq
\Psi_{\bf P}(\{{\bf p}_i\}) = N \left [ \prod_i S_i({\bf P}) \right ]
\Psi_{{\bf
P}=0}(\{\widetilde{{\bf p}}_i\})	\ \ \ .
\eeq

The model accounts for two important relativistic effects : \par

1) Lorentz contractions of the wave function
\bea
&&\widetilde{{\bf p}}_{iT} = {\bf p}_{iT} \nn \\
&&\widetilde{p}_{iz} = {E \over M} p_{iz} - {P \over M} \varepsilon_i = \sqrt{1
- \beta^2}
\left ( p_{iz} - {P \over M} m_i \right ) \nn \\
&&\widetilde{\varepsilon}_i  = {E \over M} \varepsilon_i - {P \over M} p_{iz}
\cong m_i
\eea

2) Lorentz boost of Dirac spinors

\beq
S_i({\bf P}) = \sqrt{{E + M \over 2M}} \left ( 1 + {\bfalpha_i \cdot {\bf P}
\over E + M} \right )		\
\ \ .			 \eeq

	The calculation of current matrix elements is relatively simple in the equal
velocity frame (EVF),
where the initial and final hadrons have the same modulus of the velocity
$\beta$, and opposite in
direction. An interesting relation in this frame is~:
\beq
1 - \beta^2 = {4M_iM_f\over \left ( M_i + M_f \right )^2} {1 \over \left [ 1-
{q^2 \over
\left ( M_i + M_f \right )^2} \right ]} \ \ \ . \eeq

\noi The form factors read, in this model~:
\bea
&&f_+(q2) =  {\sqrt{4M_iM_f} \over \left ( M_i + M_f \right )} {1 \over \left [
1 -
{q^2 \over \left ( M_i + M_f \right )^2} \right ]}  I(q^2)  (1 + X_+) \nn \\
&&V(q^2) =  {\sqrt{4M_iM_f} \over \left ( M_i + M_f \right )} {1 \over \left [
1- {q^2
\over \left ( M_i + M_f \right )^2} \right ]} I(q^2)  (1 + X_V) \nn \\
&&A_1(q^2) = {\sqrt{4M_iM_f} \over M_i + M_f} I(q^2)  (1 + X_1) \nn \\
&&A_2(q^2) =  {\sqrt{4M_iM_f} \over \left ( M_i + M_f \right )} {1 \over \left
[ 1 -
{q^2 \over \left ( M_i + M_f \right )^2} \right ]} I(q^2)  (1 + X_2)
\eea

\noi etc. In these expressions $I(q^2)$ is the wave function overlap, and
$X_+$, $X_V$, $X_1$, $X_2$
($M_i$, $M_f$, $q^2$) $\sim$ $O(m)$ are corrections proportional to the
spectator quark mass. \par

 	The model has a number of interesting features. It satisfies heavy-to-light
Isgur-Wise scaling
(16), (17), and also heavy-to-heavy Isgur-Wise scaling (18) with the Isgur-Wise
function
\beq
\xi (v\cdot v') = {2 \over 1 + v\cdot v'} I(v\cdot v')		\qquad \xi (1) = 1
\eeq

\noi where $I(v\cdot v')$ depends on the overlap of the wave functions at
rest~:
\beq
I(v\cdot v')  \cong  \int \Phi_{{\bf P}_f=0}^+ \left ( {\bf p} + {m \over M_i +
M_f}\widetilde{{\bf
q}} \right ) \Phi_{{\bf P}_i=0} \left ( {\bf p} - {m \over M_i + M_f}
\widetilde{{\bf q}} \right )
d{\bf p}	 \eeq

\noi where $\widetilde{{\bf q}}$ is the Lorentz contracted momentum transfer
\beq
\widetilde{{\bf q}}^2 = (1 - \beta^2) {\bf q}^2 = \left ( M_f + M_i \right )^2
{\left [ \left ( M_f - M_i \right )^2 - q^2 \right ] \over \left [ \left ( M_f
+ M_i \right )^2 - q^2
\right ]}  = \left ( M_i + M_f \right )^2 \left ( {v\cdot v' - 1 \over v\cdot
v' + 1} \right )	\ \ \
. \eeq

\noi The normalization $\xi (1) = 1$ results then from the normalization of the
wave function at rest.
In the harmonic oscillator model one has~:

\beq
\xi (v\cdot v') = {2 \over 1 + v\cdot v'}  \exp \left [ - {m^2R^2 \over
\sqrt{2}}
\left ( {v\cdot v' - 1 \over v\cdot v' + 1} \right ) \right ]	 \eeq

\noi an expression found elsewhere on different grounds$^{14}$. The slope is
given by
\beq
\rho^2 = - \xi '(1) = {1 \over 2} + {m^2R^2 \over 2\sqrt{2}}
\eeq

\noi where the second term is the dominant one in the non-relativistic limit,
and the first is a
relativistic correction due to the boost of spinors. \par

	Moreover, the model exhibits scaling corrections due to the spectator quark
mass that point to a
softening of the scaling laws (17) at fixed ${\bf q}$. These softening of the
scaling is confirmed by
lattice calculations within very large errors$^{15}$.  Also, neglecting
hyperfine splitting, the model gives
relations among the form factors that express the fact that, in the model, the
total quark spin of
the final meson equals the meson spin~:
\beq
A_0(q^2) = f_+(q^2)
\eeq
\[ 2M_f \left ( M_i - M_f \right ) f_0(q^2) = \left ( M_i^2 - M_f^2 - q^2
\right ) A_1(q^2) -
{\lambda \left ( M_i^2, M_f^2,q^2 \right ) \over \left ( M_i + M_f \right )^2}
A_2(q^2) \ \ \ . \]
\vskip 5 mm
\subsection{Example of phenomenological application}

Let us consider the relation between the semileptonic form factors $D \to
K(K^*)\ell \nu$ and the form
factors extracted from the non-leptonic decays $B \to \psi K(K^*)$. We have the
following theoretical
constraints and experimental data : \par

1) Heavy-to-light Isgur-Wise scaling near $q^2 \cong q^2_{max}$. \par

2) Data on $D \to K(K^*)\ell \nu$ near $q^2 \cong  0^{16}$~:
\bea
&&f^{sc}_+(0) = 0.77 \pm 0.04	  \qquad \qquad V^{sc}(0) = 1.16 \pm 0.16 \nn \\
&&A^{sc}_1(0) = 0.61 \pm 0.05	 \qquad \qquad A^{sc}_2(0) = 0.45 \pm 0.09
\eea

3)
Data on the decays $B \to \Psi K(K^*)$, that give information on the form
factors at $q^2 =
m^2_{\psi}$ ($L$ stands for longitudinal polarization)~:
\beq
R = {\Gamma \left ( \bar{B}_d^0 \to \psi K^{*0} \right ) \over \Gamma \left (
\bar{B}_d^0 \to \psi K^0
\right )} 	  \qquad \qquad 	R_L = {\Gamma_L \left ( \bar{B}_d^0 \to \psi K^{*0}
\right ) \over
\Gamma_{tot} \left ( \bar{B}_d^0 \to \psi K^{*0} \right )}
\eeq

\begin{center}
Table 5
\vskip 3 mm
\begin{tabular}{|l|c|c|} \hline
&$R$	&$R_L$ \\ \hline
& &0.66 $\pm$ 0.10 $\begin{array}{c} +0.10 \\ -0.08 \\ \end{array}$
(CDF)$^{21}$ \\
Experiment &1.64 $\pm$ 0.34 (CLEO)$^{23}$ &$>$ 0.78 (ARGUS)$^{22}$ \\
& &0.80 $\pm$ 0.08 $\pm$ 0.05  (CLEO)$^{22}$ \\ \hline
BSWI$^{17}$	&4.23	&0.57 \\ \hline
BSWII$^{18}$	&1.61	&0.36 \\ \hline
GISW$^{19}$	&1.71	 &0.06 \\ \hline
QCDSR$^{20}$	&7.60	&0.36 \\ \hline
\end{tabular}
\end{center}

\noi Within the factorization hypothesis$^{17}$ these rates can be related to
the $B \to K(K^*)$
form factors, like for example~:
\beq
A \left ( \bar{B}^0_d \to \psi K \right ) = {G \over \sqrt{2}}  V_{cb} V^*_{cs}
\ 2
f_{\psi} \ m_B \ f_+ (m^2_{\psi}) a_2 p
\eeq

\begin{center}
Table 6 \\
\vskip 3 mm
\begin{tabular}{|l|c|c|c|c|c|c|}	\hline
& & & & & & \\
&$R$	&$R_L$	&${f^{sc}_+(0) \over A^{sc}_1(0)}$ &${V^{sc}(0) \over
A^{sc}_1(0)}$ &${A^{sc}_2(0) \over A^{sc}_1(0)}$	&$\chi^2/DoF$ \\
& & & & & & \\ \hline
Exp. &1.64 $\pm$ 0.34	&0.66 $\pm$ 0.14	&1.26 $\pm$ 0.12	&1.90 $\pm$ 0.25	&0.74
$\pm$ 0.15 & \\
\hline
Fit &2.15 &0.45	&1.45	&1.62	&0.81	&4.2 \\ \hline
\end{tabular} \end{center}

\noi A simultaneous fit to $D$ semileptonic form factors and to $B$
non-leptonic decays is not very
good (Table 6), but some trends of our model seems to be favored, namely~: \par
1) Softened scaling~:  ${f^{sb}_+(q^2_{max}) \over f^{sc}_+(q^2_{max})}= \left
( {m_D \over m_B}
\right )^{{1 \over 2}} \left ( {m_B + m_K \over m_D + m_K} \right )$,   etc.
\par

2) Pole for the ratios $f_+(q^2)/A_1(q^2)$, $V(q^2)/A_1(q^2)$,
$A_2(q^2)/A_1(q^2)$. \par

3) Weak $q^2$ dependence of $A_1(q^2)$.

\vskip 5 mm
\subsection{The factorization problem}

In the precedent application, we have made use of factorization to establish a
connection between
non-leptonic decays and semi-leptonic form factors. For $B$ mesons,
factorization seems
ap\-pro\-xi\-ma\-te\-ly correct at least for the so-called class I decays, that
are of leading order
at large $N_c$, as shown by CLEO results$^3$. But it could happen that for the
decays $B \to \psi
K(K^*)$, that are subleading in $1/N_c$, violations of factorization (i.e. the
recipe (35)) could
modify the ratio $R$ and even $R_L$. We can wonder whether this assumption has
theoretical
justification. First, we recall that this factorization principle is
implemented in the standard
approach of SVZ$^{24}$, with particular values of $a_1$ and $a_2$, which are
found to be
\beq
a_1 = c_1 + {c_2 \over N_c} \quad , \qquad a_2 = c_2 + {c_1 \over N_c}
\eeq

\noi where $c_1$, $c_2$ are the coefficients in the effective weak
Hamiltonian~:
\beq H = c_1 (\bar{c}b)_1(\bar{s}c)_1 + c_2 (\bar{c}c)_1(\bar{s}b)_1	 \ \ \
.	 \eeq

\noi In fact it can be shown that this standard factorization is exact in the
$N_c \to \infty$
limit. However, as shown by a simple argument due to Shifman and by us$^{25}$,
the predicted
rates cannot be correct at subleading order $1/N_c$, because they contradict
duality. The argument is
as follows. Let us consider as an example the Cabibbo suppressed  process $b
\to c\bar{u}s$. There
are two types of decays corresponding to different color topologies, namely
$B_d \to D^+D_s^-$ (class
I) and $B_d \to \psi K^0$ (class II). Let us consider the effective Hamiltonian
(the subindex
means color singlet). The decay rates into a pair of hadrons corresponding to
classes I and II
result, from factorization~:
 \bea
&&\left | A \left ( B_s \to (\bar{s}c)_1(\bar{c}s)_1 \right ) \right |^2 \sim
\left [ c_1 +
{c_2 \over N_c} \right ]^2 \nn \\
&&\left | A \left ( B_s \to (\bar{c}c)_1(\bar{s}s)_1 \right ) \right |^2 \sim
\left  [ c_2 +
{c_1 \over N_c} \right ]^2	\ \ \ .
\eea

\noi Total quark decay rate differs from the sum of the expressions (38), since
it is
proportional to~:

\beq
\left | A \left ( b \to c\bar{u}s \right ) \right |^2 \sim c^2_1  +
c^2_2  + 2 {c_1 c_2 \over N_c}  \not=  \left [ c_1 + {c_2
\over N_c} \right ]^2 + \left [ c_2 + {c_1 \over N_c} \right ]^2 .\label{39}
\eeq

	On the contrary, duality tells us that by summing over all mesons one should
obtain total decay
rate $\propto  c^2_1  + c^2_2  + 2 c_1 c_2 / N_c$ . Therefore, Shifman proposes
a recipe: simply to impose duality by multiplying the subleading terms in the
decay rates to hadrons
by a factor $x = 1/2$ \bea
&&\left | A \left ( B_d \to (\bar{s}c)_1(\bar{c}s)_1 \right ) \right |^2 \sim
\left [ c_1 + x
{c_ 2 \over N_c} \right ]^2 \nn \\
&&\left | A \left ( B_d \to (\bar{c}c)_1(\bar{s}s)_1 \right ) \right |^2 \sim
\left [ c_ 2 + x
{c_1 \over N_c} \right ]^2	\ \ \ .
\eea

In our opinion, the reason for the discrepancy observed in (\ref{39}) is that
the color configurations $(\bar{c}b)_1(\bar{s}c)_1$ and
$(\bar{c}c)_1(\bar{s}b)_1$ are not orthogonal, the overlap being of order
$1/N_c$~:

\beq
(\bar{c}b)_1(\bar{s}c)_1 = {1 \over N_c} (\bar{c}c)_1(\bar{s}b)_1 +
{\sqrt{N^2_c -1} \over N_c}  \left [ (\bar{c}c)_8(\bar{s}b)_8 \right ]_1 	\ \ \
.  \eeq

We have considered a model with non-relativistic scalar quarks bound to color
singlets by a color
harmonic oscillator potential. This model takes into account of a Final State
Interaction that
automatically restores duality or the conservation of probability. Indeed,
summing over all mesons in
the limit in which the radius $R \to \infty$ one should find the free quark
result. This is indeed the
case, since one finds~:
\bea
&&\left | A \left ( B_d \to (\bar{s}c)_1(\bar{c}s)_1 \right ) \right |^2 \sim
\left [ c_1 + y
{c_2 \over N_c} \right ]^2 \nn \\
&&\left | A \left ( B_d \to (\bar{c}c)_1(\bar{s}s)_1 \right ) \right |^2 \sim
\left [ c_2 + (1
- y) {c_1 \over N_c} \right ]^2
\eea

\noi with $y(m_c,m_s)$ depending non-trivially on the
masses, unlike the universal factor 1/2 proposed by Shifman. It is seen that
FSI restores automatically duality. Notice by the way that
expression (42), for an arbitrary value of $y$, is more general than Shifman's
Ansatz and would as well restore duality. Another study of restoration of
duality in the Quark Model approach may be found in [26]. The ratios $R$ and
$R_L$ could be modified by this type of FSI effects, since by introducing
spin there is no reason a priori that the ratio $R_L$ would not be modified
also.

\section*{Acknowledgements}

We would like to thank the Clermont-Ferrand team who has so much contributed to
the success of this meeting.
This work was supported in part by the CEC Science Project SC1-CT91-0729 and by
the Human Capital and Mobility Programme, contract CHRX-CT93-0132.

\noi


\begin{thebibliography}{11}

\bibitem{} A. Le Yaouanc, L. Oliver, O. P\`ene and J.-C. Raynal, Hadron
Transitions in the Quark
Model, Gordon and Breach (1988). See also the recent comparison between the
Quark Model and QCD by K.
G. Wilson and D. G. Robertson, preprint OSU-NT-94-08.
\bibitem{} K. G. Wilson, Phys. Rev. {\bf D10},
2445 (1974).
\bibitem{} K. D. Born {\it et al.}, Nucl. Phys. B (Proc. Suppl.) {\bf 20}, 394
(1991).
\bibitem{} E. Eichten and F. Feinberg, Phys. Rev. {\bf D23}, 2724 (1981)~; W.
Lucha, F. F. Sch\"oberl
and D. Gromes, Phys. Rep. {\bf 200}, 127 (1991)~; N. Brambilla, P. Consoli and
G. M. Prosperi, Phys.
Rev. {\bf D50}, 5878 (1994).
 \bibitem{} See, for example, A. Le Yaouanc {\it et al.}, Phys. Rev. {\bf D29},
1223 (1984)~; {\bf
D31}, 137 (1985)~; V. Bernard, A. A. Osipov and U. G. Meissner, Phys. Lett.
{\bf B285}, 119 ().
\bibitem{} Cornell model levels as quoted in [7].
\bibitem{} A. Barchielli, N. Brambilla and G. M. Prosperi, Nuovo Cimento {\bf
103A}, 59 (1990).

\bibitem{} R. Mac Lary and N. Byers, Phys. Rev. {\bf D28}, 1692 (1983). For the
$c\bar{c}$ system, the model gives E1 transitions $\chi_c(1^3P_J) \to J/\psi$
somewhat too large.
\bibitem{} N. Isgur and M. B. Wise, Phys. Lett. {\bf 232B}, 113 (1989)~; {\bf
237B}, 527 (1990).
\bibitem{} N. Isgur and M. B. Wise, Phys. Rev. {\bf D42}, 2388 (1990).

\bibitem{} M. Neubert, CERN-TH.7395/94, to appear in Physics Letters, and talk
at the International
Conference on High Energy Physics, Glasgow, Scotland (1994), CERN-TH.7396/94.
 \bibitem{} T. Bowder, to appear in Proceedings of the Cornell preprint CLNS
94/1285 (1994).
  \bibitem{} See, for example, R. Aleksan, A. Le Yaouanc, L. Oliver, O. P\`ene
and J.-C. Raynal,
preprint DAPNIA/SSP 94-24, LPTHE 94/15, to appear in Phys. Rev. D.
\bibitem{} M. Neubert and V. Rieckert, Nucl. Phys. {\bf B382}, 97 (1992).
\bibitem{} As Abada {\it et al.}, Nucl. Phys. {\bf B416}, 675 (1994); A. Le
Yaouanc and O. P\`ene, Third Workshop on the
Tau-Charm Factory (1993); C.R. Allton et al. Phys. Lett {bf 345}, 513 (1995).
\bibitem{} M. S. Whitherell, talk at International Symposium on Lepton and
Photon Interactions at High Energies, Cornell, Ithaca, N. Y., UCSB-HEP-93-06.
\bibitem{} M. Wirbel, B. Stech and M. Bauer, Z. Phys. {\bf C29}, 637 (1985)~;
{\bf C34}, 103 (1987).
\bibitem{} M. Neubert {\it at al.}, in {\it Heavy Flavors}, eds. A. J. Buras
and M. Lindner, World
Scientific, Singapore (1992).
\bibitem{} N. Isgur, D. Scora, B. Grinstein and M. B. Wise, Phys. Rev. {\bf
D39}, 799 (1989).
\bibitem{} P. Ball, Phys. Rev. {\bf D48}, 3190 (1993).
\bibitem{} F. Abe et al., Fermilab-Conf-93-200-E, 16th International Symposium
on Lepton and Photon
Interactions, Ithaca, N.Y. (1993).
 \bibitem{} M. Danilov, talk given at the ECFA Working Group on $B$ Physics,
DESY (1992).
\bibitem{}  M. S. Alam et al. (CLEO collaboration), Phys. Rev. {\bf D50}, 43
(1994)~; A. J. Buras,
preprint MPI-PhT/94-60.
\bibitem{} A. I. Vainshtein, V. I. Zakharov and M. A. Shifman, JETP {\bf 45},
670 (1977).
\bibitem{} M. A. Shifman, Nucl. Phys. {\bf B388}, 346 (1992)~; R. Aleksan, A.
Le Yaouanc, L. Oliver,
O. P\`ene and J.-C. Raynal, Phys. Lett. {\bf B316}, 567 (1993).
\bibitem{} A. Le Yaouanc, L. Oliver,
O. P\`ene and J.-C. Raynal, LPTHE-Orsay 95/26.
 \end{thebibliography}
\end{document}